\documentclass[final,5p,times,twocolumn]{elsarticle}
\usepackage{amssymb}
\usepackage{lipsum}
\usepackage{amsthm}
\usepackage{mathptmx}
\usepackage{tikz}
\usepackage{tikz-cd}
\usepackage{mathtools}
\DeclarePairedDelimiter\bra{\langle}{\rvert}
\DeclarePairedDelimiter\ket{\lvert}{\rangle}
\DeclarePairedDelimiterX\braket[2]{\langle}{\rangle}{#1 \delimsize\vert #2}
\journal{Physics Letters B}
\begin{document}
\begin{frontmatter}
\title{Primordial Black Holes and Instantons: Shadow of an Extra Dimension}
\author{Reinoud Jan Slagter\corref{cor1}\fnref{label2}}
\ead{info@asfyon.com/reinoudjan@gmail.com}
\ead[url]{www.asfyon.com}
\fntext[label2]{Asfyon, Astronomisch Fysisch Onderzoek Nederland}
\cortext[cor1]{Former: University of Amsterdam}
\begin{abstract}
We investigated an exact solution in a conformal invariant Randall-Sundrum 5D warped brane world model on a time dependent Kerr-like spacetime. The singular points are determined by a quintic polynomial in the complex plane and fulfills Cauchy's theorem on holomorphic functions. The solution, which is determined by a first-degree differential equation, shows many similarities with an instanton. In order to describe the quantum mechanical aspects of the black hole solution, we apply the antipodal boundary condition. The solution  is invariant under time reversal and also valid in Riemannian space. Moreover, CPT  invariance in maintained. The vacuum instanton solution follows from the 5D as well as the effective 4D brane equations, only when we allow the contribution of the  projected 5D Weyl tensor on the brane (the KK-'particles'). The topology of the effective 4D space of the brane is the projective $\mathbb{R}P^3$ (elliptic space) by identifying antipodal points on $S^3$. 
The 5D is completed by applying the Klein bottle embedding and the 
$\mathbb{Z}_2$ symmetry of the RS model.
This model fits very well with the description of the Hawking radiation, which remains pure. 
We have also indicated a possible way to include  fermions. Our 5D space admits a double cover of $S^3$ and after fibering to the $S^2$, we obtain the effective  black hole horizon. The connection with the icosahedron discrete symmetry group is investigated. It seem that Bekenstein's conjecture that the area of a black hole is quantized, could be applied to our model.
\end{abstract}
\end{frontmatter}
\section{Introduction}\label{1}
Following the pioneering work of Hawking \cite{hawking1975} on the evaporation of black holes, many scientists began to study the quantum-gravitational properties of black holes. However, there are some profound unsolved problems as one approaches the Planck scale. Due to the extremely high curvature, quantum effects can no longer be denied. It is conjectured that knowledge of the quantum effects near the horizon of a black hole will eventually lead to a consistent quantum gravity model. 
The properties of the black hole, i.e. its entropy, mass and the emission of radiation, need to be revised because there are some serious paradoxes. 
These paradoxes involve the problems that arise with the quantum effects that occur at the horizon. In particular, the information-, complementarity- and firewall paradoxes. Furthermore,  there is the issue of unitarity of the S-matrix.
Another amazing conclusion on black hole physics concerns the amount of  information that can be stored in a black hole. It is proportional to the  area ($A$) of the sphere. It is expressed by the famous  Bekenstein-Hawking (BH) entropy of a black hole \cite{bekenstein1972}, i.e. $S_{BH}=k_B c^3 A/4G_N\hslash$. 
This phenomenon can be related to the holographic principle, first mentioned by 't Hooft \cite{thooft1993}. A famous  application of this principle is the AdS/CFT correspondence.   It refers to the astonishing dualities between field theories with and without gravitation. It is also  referred to as the gauge
theory-gravity correspondence. There is a relationship between $d$-dimensional gravitation and a local field theory without gravitation in $(d-1)$ dimensions. Famous is the equivalence between type IIB string theory on $AdS_5\times S^5$ and four-dimensional $N=4$ super Yang-Mills (YM) theory \cite{mal1999,boer2015}.
An other well known concept is the adiabatic invariant of black holes which can be used  to derive the quantized entropy spectrum of the black hole. Even the horizon area could be an adiabatic invariant using the quasi-normal modes. Bekenstein claimed already 50 years ago \cite{bekenstein1974}, that a black hole  in the GRT plays the same role as an atom in the emerging QM. Would a black hole therefore also have a discrete emission spectrum \cite{bekenstein2002,thooft2019} and 
perhaps  observable?
But then also the surface area. The surface area appears again in the  entropy. So the entropy should exhibit the same discrete spectrum.
However, one must keep in mind that a rigorous quantum discussion of black holes must await the development of a complete quantum theory of gravitation. Nevertheless, it is possible to consider heuristically some quantum features of black holes by means of well-understood principles and arguments from ordinary quantum theory.
It is then a natural step to consider the principle in a more general context.
We will use the Randall-Sundrum (RS) \cite{ran1999,ran1999a}  five dimensional  warped brane-world model.  The model is a simplification of the full string model. The extra-dimensional bulk space has a negative cosmological constant. The dilution of gravity by the presence of the bulk effectively weakens gravity on the brane. The result is that the true higher dimensional Planck scale can be much lower than the effective 4D Planck scale, possibly down to the electroweak scale. It also extends the range of graviton modes that can be felt on the brane, the so-called Kaluza-Klein (KK) modes. The model does not rely on compactification to localize gravity on the brane, but on the curvature of the bulk. One calls it a 'warped' compactification.
The most simple model is given by the metric
$ds_{RS}^2=e^{-2|y_5|/y_0}\eta_{\mu\nu}dx^\mu dx^\nu +dy_5^2$, with $y_0$ the curvature radius of the $AdS_5$, related the the bulk cosmological constant and  $y_5$ the bulk coordinate.
Some physicists believe that a partial solution of the quantum gravity issue, can be found in topology.
One can apply Poincar\'e's theorem, i.e.
{\it every three-dimensional topological manifold which is closed, connected and has a trivial fundamental symmetry group, is homeomorphic to the three-sphere $S^3$}.
It came out that the theorem is only valid for the 3-sphere, by using complex coordinates.
The proof was based on the group  of rotational symmetries of the icosahedron, isomorphic to the alternating group $A_5$. 
We conjecture for our new solution for the Kerr-like black hole that 
{\it the adiabatic invariants of the area of the black hole, resulting in a discrete spectrum,  will be related to the discrete representation of the zero's of the quintic in the 5D conformal invariant vacuum  black hole solution in the Randall-Sundrum model.} 
We need, however, a conformal invariance (CI) Lagrangian. Local conformal invariance is then spontaneously broken, comparable with the Higgs mechanism in the Standard Model (SM), which has no intrinsic mass scale without the Higgs field. It would be natural if gravity also had no length or mass scale. CI could ensure this. The conformal transformations on the metric and scalar field are 
\begin{equation}
g_{\mu\nu}\rightarrow \Omega({\bf x})^{4/(d-2)}g_{\mu\nu}, \quad\Phi\rightarrow \Omega({\bf x})^{(2-d)/2}\Phi\label{1-1}
\end{equation}
The scalar field can be accompanied by a dilaton field $\omega$, by writing $g_{\mu\nu}=\omega^{4/(d-2)} \tilde g_{\mu\nu}$ \cite{thooft2014,thooft2015}.
Moreover, one should like to consider  Riemannian geometry in stead of pseudo-Riemannian spacetimes. This can be done by the Wick rotation $\omega\rightarrow i\omega$ together with $t\rightarrow i\tau$. The reason is twofold. First, quantum computations can be easily done and secondly, on very small scales, the dilaton can be treated on equal footing as the quantum scalar field. $\tilde g_{\mu\nu}$ can then we treated 'classically'. 
The basis of our model relies on the elliptic interpretation of the manifold, already introduced by Schr\"odinger \cite{schrod1957,gibbons1986,thooft2016}.
The topology of the 5D RS model is written as $S^3\times \mathbb{R}$ and we apply  the antipodal identification on the 3-sphere my means of the stereographic projection $P\mathbb{R}^3\sim S^3/\mathbb{Z}_2$, together with the Hopf fibration. 
Further, the use of the Klein bottle ($\mathbb{K}^4$) structure fits very well in the RS model, because it can be embedded in $\mathbb{R}^4$ without obstruction. 
Further, it is the direct sum of two projected planes, $\sim \mathbb{R}P^2\#\mathbb{R}P^2$.

The structure of this letter is as follows. In section two we present the new model. In section three we summarized the new topology described by the embedding of the Klein surface in the 5D model  and the related Hopf fibrations.
In section four we make the connection with the instanton and  a possible treatment of the Hawking particles.
\section{The model}\label{2}
Recently , we studied a model which relies on the RS model in a conformal invariant  dilaton-gravity theory \cite{slagter2023,slagter2024,slagter2025,slagter2025b}, on the spacetime,
\begin{eqnarray}
ds^2=\omega(t,r,y_5)^{4/3}y_0\Bigl[-N(t,r)^2dt^2+\frac{1}{N(t,r)^2}dr^2+dz^2\cr +r^2(d\varphi+N^\varphi(t,r)dt)^2+dy_5^2\Bigr]\qquad\quad \label{2-1}
\end{eqnarray}
where $y_5$ is the fifth coordinate, $y_0$ the bulk dimension and $\omega$ is a warp factor, reinterpreted as a dilaton field. We write $^{(5)}{g_{\mu\nu}}=\omega^{4/(d-2)} {^{(5)}{\tilde g_{\mu\nu}}}$ and $\tilde g_{\mu\nu}=^{(4)}{\tilde g_{\mu\nu}}+n_\mu n_\nu$. An 'unphysical' spacetime is thus  separated. Here $n_\mu$ is the unit normal to the brane. Again we write $^{(4)}{\tilde g_{\mu\nu}}=\bar\omega^2{^{(4)}{\bar g_{\mu\nu}}}$.
The Lagranian under consideration is
\begin{eqnarray}
S=\int d^dx\sqrt{-\tilde g}\Bigl[-\frac{1}{2}\xi (\Phi\Phi^*+\omega^2)\tilde R\cr-\frac{1}{2}\tilde g^{\mu\nu} \Bigl({\cal D}_\mu\Phi({\cal D}_\nu\Phi)^*+\partial_\mu\omega\partial_\nu\omega\Bigr)\cr
-\frac{1}{4}F_{\mu\nu}F^{\mu\nu}-V(\Phi,\omega)-\Lambda\kappa^{\frac{4}{d-2}}\xi^{\frac{d}{d-2}}\omega^{\frac{2d}{d-2}}\Bigr]\label{2-2}
\end{eqnarray}
where ${\cal D}$ is the gauge covariant derivative and $F$  the Abelian field strength. Furthermore, $\xi=(d-2)/(4(d-1)$ and we applied the Wick rotations $\omega^2\rightarrow -6\frac{\omega^2}{\kappa^2}$, with $\kappa=8\pi G_N$, to ensure that the field $\omega$ has the same unitarity and positivity properties as the scalar field. The model is invariant under  Eq. (\ref{1-1}).
The $\omega$ freedom is used by the in-falling observer to describe his experience of the vacuum.
Since we are working in a RS warped brane world model, one solves the 5D and the effective 4D equations simultaneously, with the contribution of the projected bulk Weyl tensor appearing in the latter.
The 4D and 5D equations must be solved simultaneously, i. e.
\begin{equation}
{^{(5)}}{G_{\mu\nu}}=-\Lambda_5{^{(5)}g_{\mu\nu}}\label{2-3}
\end{equation}
\begin{equation}
{^{(4)}G_{\mu\nu}}=-\Lambda_{eff}{^{(4)}g_{\mu\nu}}+T_{\mu\nu}^{(\omega)}-{\cal E}_{\mu\nu}\label{2-4}
\end{equation}
where $T_{\mu\nu}^{(\omega)}$  is the contribution from the dilaton (which also appears in  5D form in the 5D equations), because we use the 'un-physical' metric, 
\begin{equation}
T_{\mu\nu}^\omega=\tilde\nabla_\mu\tilde\nabla_\nu\omega^2-\tilde g_{\mu\nu}\tilde\nabla^2\omega^2+\frac{1}{\xi}\Bigl(\frac{1}{2}\tilde g_{\alpha\beta}\tilde g_{\mu\nu}-\tilde g_{\mu\alpha}\tilde g_{\nu\beta}\Bigr)\partial^\alpha\omega\partial^\beta\omega
\end{equation}
It is the contribution  ${\cal E}_{\mu\nu}$ from the projected Weyl tensor from the  bulk on the brane which makes the system consistent.
In conform invariant form, they become respectively,
\begin{equation}
\omega^2{^{(5)}}{G_{\mu\nu}}-{^{(5)}}{T_{\mu\nu}}^{(\omega)}
+\frac{3}{4\sqrt[3]{12}}\Lambda_5\kappa_5^{4/3}\omega^{10/3}{^{(5)}}{g_{\mu\nu}}=0\label{2-5}
\end{equation}
\begin{equation}
{^{(4)}}{G_{\mu\nu}}-\frac{1}{\omega^2}\Bigl[{^{(4)}}{T_{\mu\nu}}^{(\omega)}-\frac{1}{6}\Lambda_{eff}\kappa_4^2\omega^4{^{(4)}}{g_{\mu\nu}}\Bigr]+{\cal E}_{\mu\nu}=0\label{2-6}
\end{equation}
From these field equations, we obtain the PDE's,
\begin{equation}
\ddot\omega=-N^4\omega''+\frac{n}{\omega(n-2)}\Bigl(N^4\omega'^2+\dot\omega^2\Bigr)\label{2-7}
\end{equation}
\begin{eqnarray}
\ddot N=\frac{3\dot N^2}{N}-N^4\Bigl(N''+\frac{3N'}{r}+\frac{N'^2}{N}\Bigr)\qquad\qquad\cr\hspace{-0.3cm}
-\frac{n-1}{(n-3)\omega}\Bigl[N^5\Bigl(\omega''+\frac{\omega'}{r}+\frac{n}{2-n}\frac{{\omega'}^2}{\omega}\Bigl)+N^4\omega' N'+\dot\omega\dot N\Bigr]\label{2-8}
\end{eqnarray}
The values for $n$ are 4 and 5 for the 4D and 5D solutions respectively.
The solutions are\footnote{On a FLRW spacetime, the dilaton field is then considered as a warp factor, which can also be expressed exactly in $(r,t,y_5)$, where the bulk part becomes $e^{\sqrt{\frac{\Lambda_5}{6}(y_5-y_0)}}$ (\cite{slagpan2016}). You will need these solutions when the Hawking particles become hard.}
\begin{equation}
N^2=\frac{C_2}{r^2(t -b)^4+C_3}\Bigl[\frac{(r-a)^{k+1}\Bigl(r(k+1)+a\Bigr)}{k+2}+C_1 \Bigr]\label{2-9}
\end{equation}
\begin{eqnarray}
\omega = \Bigl(\frac{c}{(r-a)(t-b)}\Bigr)^{\frac{1}{2}n-1},N^\varphi=\int \frac{dr}{r^3\omega^{\frac{n-1}{n-3}}}+F_n(t)\label{2-10}
\end{eqnarray}
with constants $(a,b,c,C_i)$. Furthermore, $F_n(t)$ is a arbitrary function determined by a constraint equation. The equations are invariant for $t\rightarrow -t$.
The value $k=3$ determines the solution for our situation.  The solution is separable and we write $N^2=N_1(r)^2N_2(t)^2$. The r-dependent part fulfills the first order equation
\begin{equation}
rN_1\frac{\partial N_1}{\partial r}+N_1^2=\frac{k+1}{2}(r-a)^k\label{2-11}
\end{equation}
In general, the metric component $N_1$ is quintic polynomial, with in general five zero's.
For $C_1=0$ the zeros are at $r_H=a$ with multiplicity $(k+1)$ and at $r_H=-\frac{a}{k+1}$. 
For $k=3$ and varying $C_1$, one obtains a distribution of the zero's which are
related to  the stereographic projection of the icosahedron.  The general solution of the quintic can be found by means of elliptic curves \cite{bartlett2024}.
For $k=2$ and $C_1 < a^4/4$ there are 4 zeros, as in the BTZ solution, two of them negative. 
For $k=1$ and $C_1 < -a^3/3$ there is still a real solution.
For $k=0$ we  surprisingly get two horizons $r_H=\pm a$. This solution is also found from Eq. (\ref{2-8}) for constant $\omega$. The differential equation for $N$ becomes $N''+3N'/r+N'^2/N=0$, with solution $N\sim(1-a^2/r^2)$.

Because we have to solve the 5D and 4D equations simultaneously, we get two constraint equations with the cosmological constants $\Lambda_5$ and $\Lambda_{eff}$. By eliminating the $r$-dependent part. one obtains a 'fine tuning', 
\begin{equation}
\Lambda_5=\frac{2\sqrt[3]{12}c^{2/3}\kappa_4^2}{9\kappa_5^{4/3}}\Lambda_{eff}\label{2-12}
\end{equation}
This proves once again that the 5D and 4D effective equations must be solved together, and that the cosmological constants can be fine tuned. The situations changes, when we incorporate a scalar field. 
Cauchy's theorem on holomorphic functions $F(z)$ on an open set of the complex plane, tells us that within a simply-connected region with no singularities inside a loop C, that
\begin{equation}
\oint_C F(z)dz=0\label{2-13}
\end{equation}
Our solution for $N_1$ can also be written as
\begin{equation}
N_1^2=\frac{4}{z^2}\int z(z-a)^3dz=0\label{2-14}
\end{equation}
So we have $F(z)=z(z-a)^3$.
One applies  Cauchy's integral theorem
\begin{equation}
F(z)=\frac{1}{2\pi i}\oint_C\frac{F(\xi)}{\xi -z} d\xi
\end{equation}
with z inside C. So the values of an analytic function are determined by the values on the boundary.
This is also true for the higher derivatives in our situation, i.e. for  $F'=(z-a)(z-a/4), F''=(z-a)(z-a/2)$ and $ F'''=(z-3a/4)$.
It turns out that $z=(0, a)$ are removable singularities; they lie on the real axis and one calculates  Cauchy's principal values.
By the antipodal map, there are no fixed points, so the quintic determines the singular points on $S^3$ inside C and delivers no problems when C shrinks to a point. No crossing with the singular points occurs, due to the extra bulk dimension.
We  still have the $\Omega$ freedom of Eq. (\ref{1-1}) in order to make  the outside observer experience  a Kerr-like black hole.
The effective 4D spacetime  in suitable coordinates becomes
\begin{equation}
ds_4^2=\omega^{4/3}\bar\omega^2\Bigl[\frac{N_1^2}{N_2^2}(-dt^{*2}+dr^{*2})+dz^2+r^2d\varphi^{*2}\Bigr]\label{2-15}
\end{equation}
with
\begin{equation}
r^*=\frac{1}{4}\sum_{r^H_i}\frac{r^H_i \log(r-r^H_i)}{(r^H_i+a)^3}, t^*=\frac{1}{4C_2}\sum_{t^H_i}\frac{\log(t-t^H_i)}{(t^H_i+b)^3}\label{2-16}.
\end{equation}
The sum it taken over the roots of $(10a^3r^2+20a^2r^3+15ar^4+4r^5+C_1)$ and $C_2(t+b)^4+C_3$, i. e. $r^H_i$ and $t^H_i$.  We also defined the azimuthal angle coordinate $d\varphi^*=d\varphi+(N^\varphi/N_2^2) dt^{*}$, which can be used when an incoming null geodesic falls on the horizon, i.e. in a coordinate system that rotates about the z-axis relative to the Boyer-Lindquist coordinates. We are mainly interested in  the solution for a local observer.
In Kruskal–Szekeres  lightcone coordinates $(U,V)$ we have
\begin{eqnarray}
ds^2=\omega^{4/3}\bar\omega^2\Bigl[\frac{N_1^2}{N_2^2}\frac{dUdV}{\epsilon^2 UV}+dz^2+r^2d\varphi^{*2}\Bigr]\cr
=\omega^{4/3}\bar\omega^2\Bigl[H(\tilde U,\tilde V)(d\tilde U d\tilde V +dz^2+r^2d{\varphi^*}^2\label{2-17}
\end{eqnarray}
with $\epsilon$ a constant related to the surface gravity at the horizon and  $\tilde U=\tanh U, \tilde V=\tanh V$.
\begin{figure}[h]
	\centerline{
	\fbox{\includegraphics[width=6.5cm]{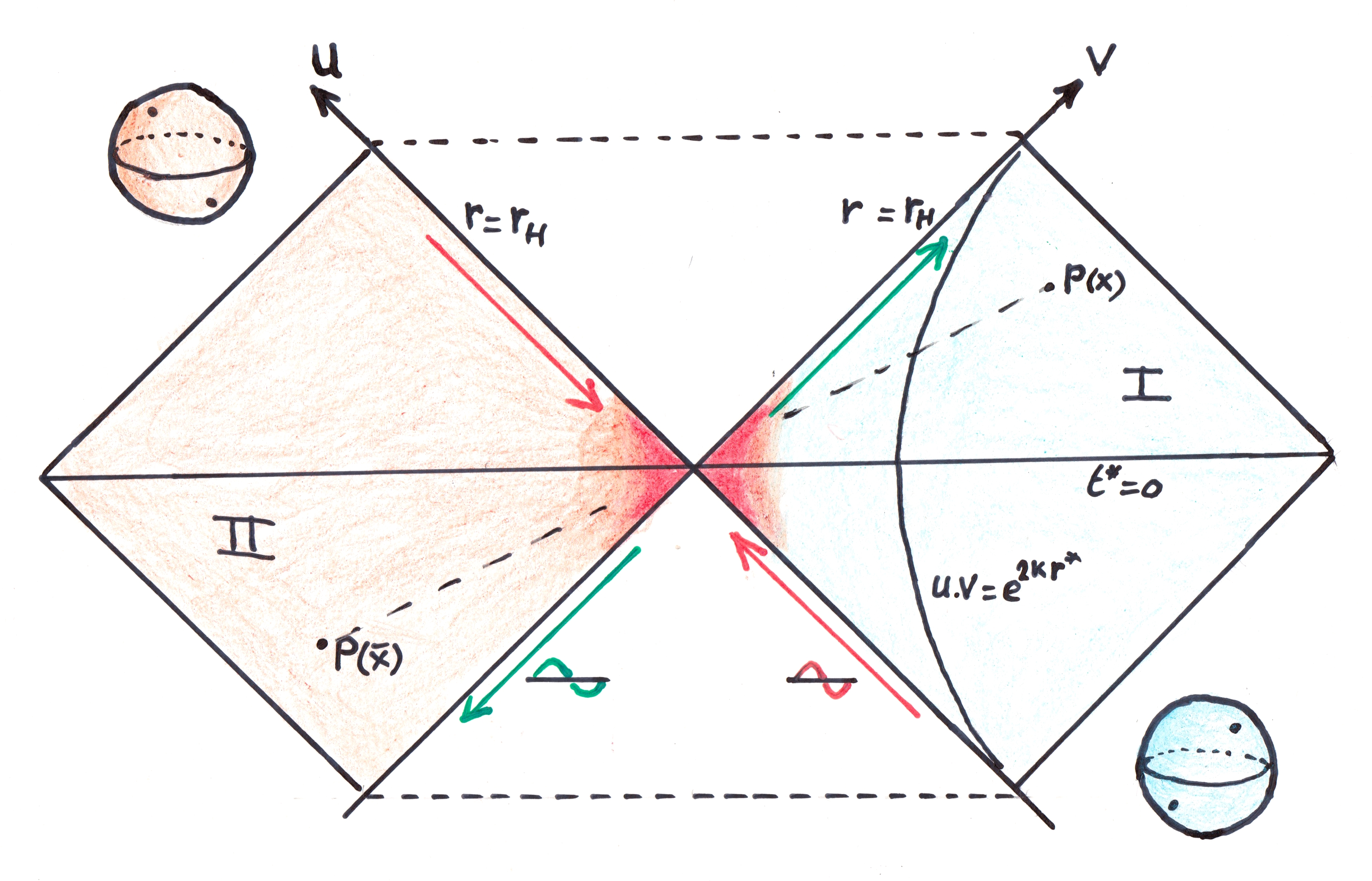}}}	
	\caption{{\it Penrose diagram for our 5D model, stereographic projected from $S^3$. The antipodal points $P(X)$ and $P(\bar X)$ are identified. Particles entering the black hole will create waves that approach the horizon from the outside, and those that pass the horizon will emerge from the 'other side' of the black hole. Note that $t^*=log(U/V)\sim log(t-t_H)$. In this approach, I and II are CPT invariant. This means that time runs backwards in II.}} 
	\label{penrose}
\end{figure}
If we omit the contribution of the projected Weyl tensor in the effective 4D Einstein equations, we obtain the Ba\u nados-Teitelboim-Zanelli 3D solution \cite{banadoz1992,compere2018}. This is equivalent to the substitution $k=2$ in Eq. (\ref{2-9}).
We also found a remarkable connection with the icosahedral M\"obius group $A_5$ when inscribed in $S^2$. The isometry group is of order 5 and has three orbits which are invariant under the antipodal map.
Our general polynomial of $N^2$ is a quintic, which  cannot be a coincident  \cite{slagter2023}. 
If we include a complex scalar-gauge $(\Phi,A_\mu)$, we find from the superfluous dilaton equation, that the scalar-dilaton potential $V(\omega,\Phi)$ satisfies
\begin{equation}
V(\omega,\Phi)=\beta_1 \Phi^{\beta_2}\omega^{\frac{2}{3}-\frac{\eta\beta_2}{y_0}}\label{2-18}
\end{equation}
with $(\beta_1,\beta_2)$ some constants. So we get a quartic conformal invariant matter coupling $\sim \Phi^2\omega^2$ for $\beta_2=2, |y_0|=3/2\eta$, where $\eta$ is the vacuum expectation. We have used in this case, a constant gauge field.
\subsection{Some conjectures}\label{2-3}
Before we proceed, we formulate some conjectures.\\
{\bf 1.} {\it  The zeros of the fifth degree polynomial, an exact solution of a black hole in a conformally invariant 5D warped spacetime, lie in the stereographically projected complex plane of the Riemann sphere around an icosahedron.  During  the evaporation of the black hole by Hawking radiation, the zero points undergo a transformation, because  the Riemann sphere shrinks. The transformation complies with the symmetry group of the icosahedron and shows similarities with the flexibility of the construction of the icosahedron.
The final shape arises from the special case $c=a^5$, on both sides of $r=0$, i.e. $a$ with multiplicity 4, and  $-a/4$. This means in the Penrose diagram for the black hole, that the central singularity is absent.
Using Cauchy's theorems and the antipodicity, one proves that there are no essential singularities.
In general, during the evaporation, the five zero points will tend to a symmetrical distribution. }\\

\begin{figure}[h]
\centerline{
\fbox{\includegraphics[width=3.35cm]{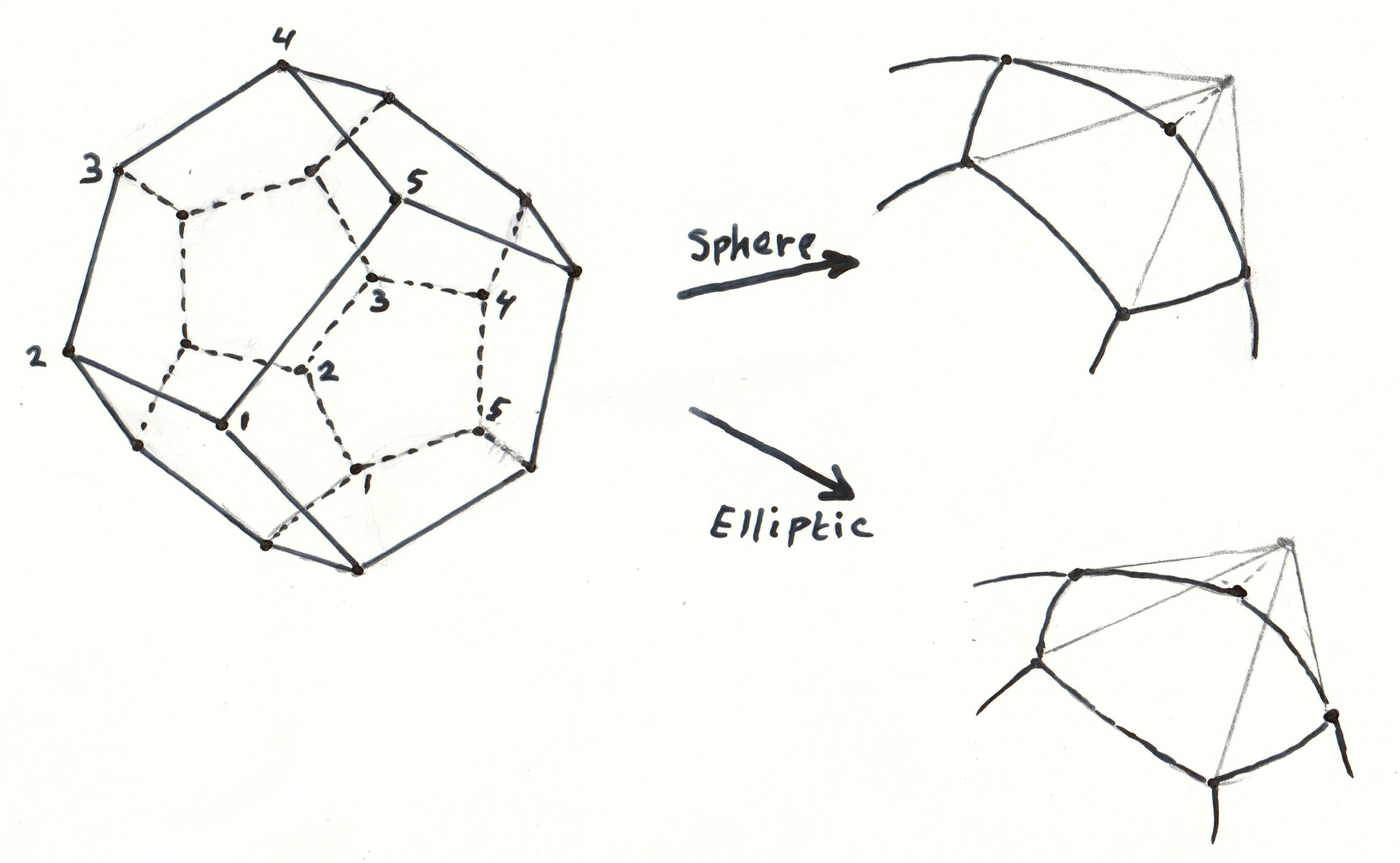}}
\fbox{\includegraphics[width=4.64cm]{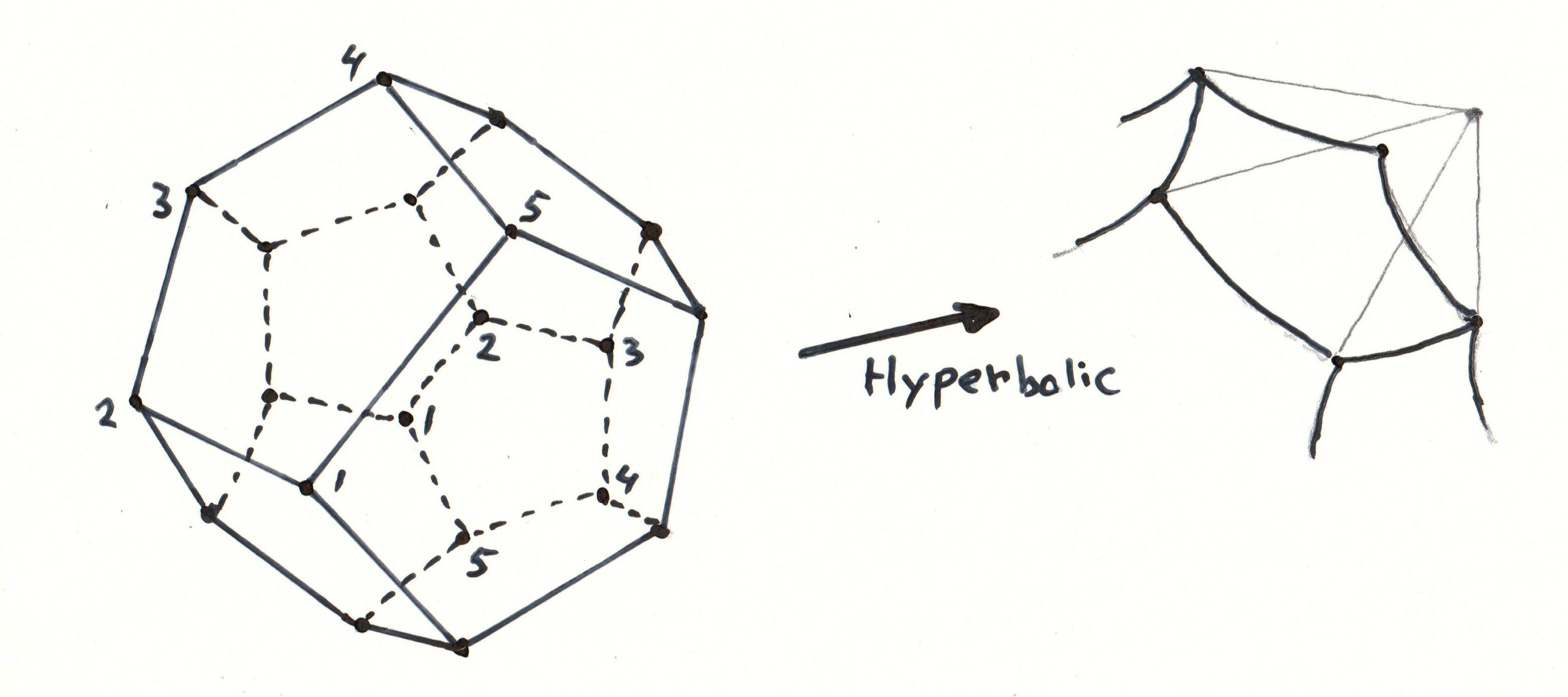}}}
\caption{{\it  The two different glued  Poincar\'e dodecahedron spaces. Left the spherical situation (and elliptic) and right the hyperbolic (Seifert-Weber space).}}\label{poincare}
\end{figure}

{\bf 2.} {\it  We discussed the properties of the three sphere in relation to the inscribed dodecahedron and icosahedron. The five vertices can now represent the zero points of our polynomial $N(r)^2$. It turns out  that the geometric structure of $S^3$ can change from spherical Euclidean to elliptic (by identifying antipodal points) and to hyperbolic (Fig.(\ref{poincare})), by changing the dimension of the dodecahedron. Our differential equations were elliptic and became hyperbolic (Riemann space) by switching from $t$ to $i\tau$. This is therefore equivalent to allowing the antipodicity. The change in the topology that we described with the three-sphere and the dodecahedron yields the transition from 5 zero points to 2, one of which is a horizon. This change is triggered by Hawking radiation. The two parameters in our polynomial must therefore also be related to the mass and a second 'charge',  related to the topology. This second parameter is not related to the angular momentum. The angular momentum  is given by $N^\varphi$. }\\

{\bf 3.} {\it The special case $c=a^5$ in the quintic, which can be written as $-r^5+5r^4(r-a)-10r^3(r-a)^2+10r^2(r-a)^3+c=0$, represents an instanton.}

{\bf 4.} {\it We conjecture that the zero's of our model can discretely change during the formation and evaporation of the black hole. This conjecture, as we will see, is based on the description of the zero's by means of the torsion points on elliptic curves. For some special values of the parameter c, we found the five roots lying in almost symmetrical regular pentagon. Using the elliptical curve method, one could find this configuration. The last non-regular situation could result in our instanton configuration. This discretely feature could be seen as a kind of a hydrogen atom, one old idea of Bekenstein.}\\
\begin{figure}[h]
	\centerline{
	\fbox{\includegraphics[width=3.cm]{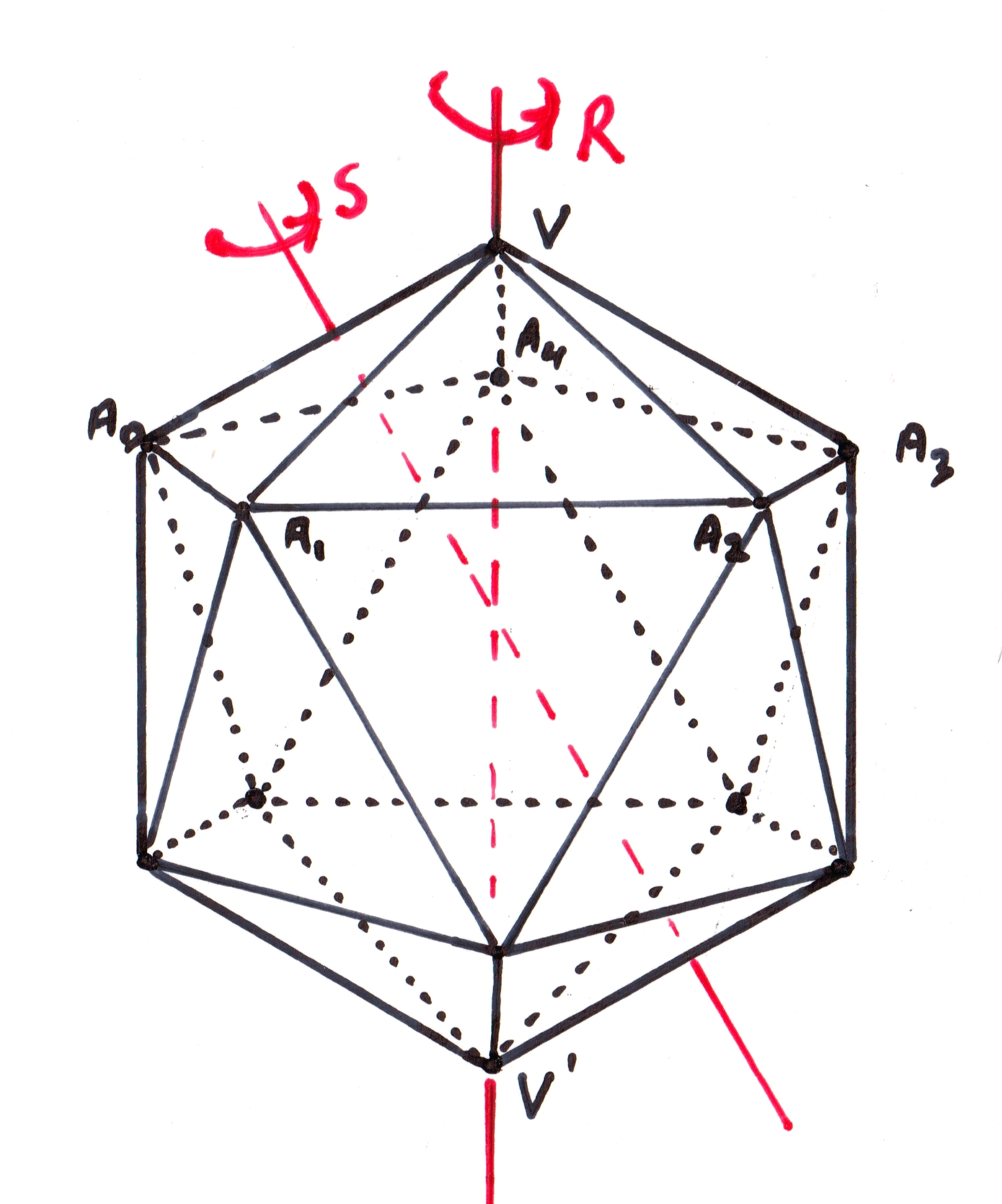}}
	\fbox{\includegraphics[width=4.38cm]{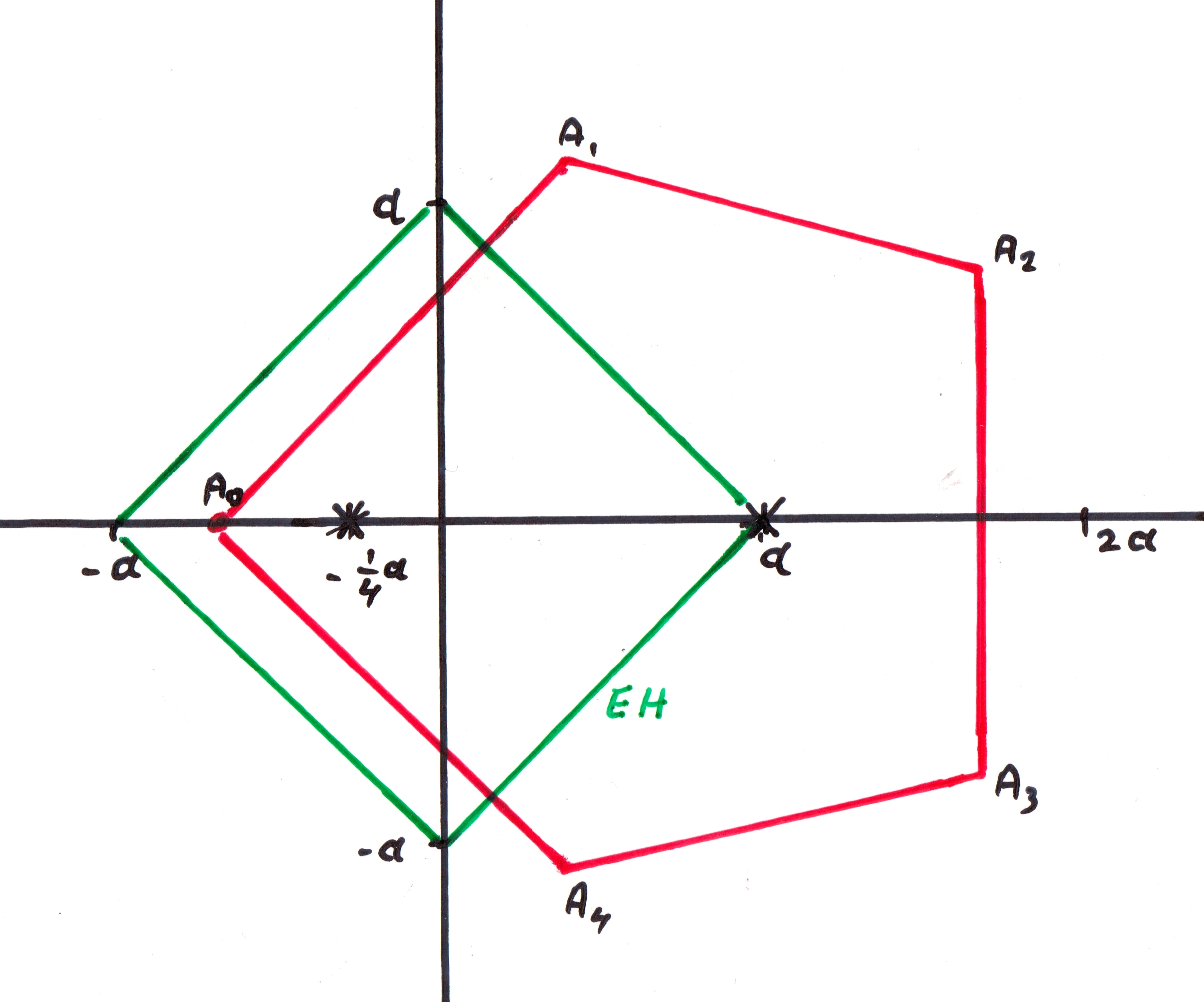}}}	
	\caption{{\it  Left: The icosahedron. It has 20  faces, 12 vertices, and  30 edges. There are 60 rotations that leave the figure invariant. We have rotations through $2\pi/5$ over $VV'$ (R) and rotations through $\pi$ (S).
Right: The five zeros for the two cases $c=a^5$ (star) and  $c=\frac{3381}{256}a^5 (A_i)$). In green  the four zero points for the self-dual Eguchi-Hanson (EH) instanton solution.}} 
	\label{elliptic}
\end{figure}
\begin{figure}[h]
	\centerline{
	\fbox{\includegraphics[width=2.5cm]{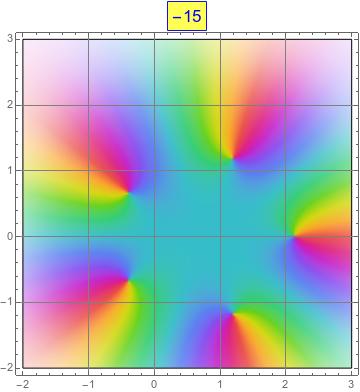}}
	\fbox{\includegraphics[width=2.5cm]{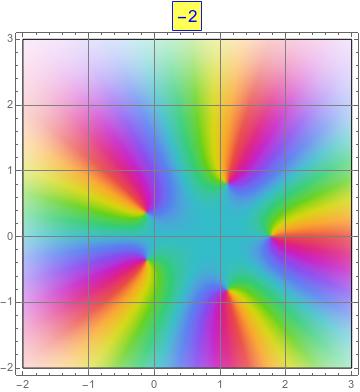}}
	\fbox{\includegraphics[width=2.5cm]{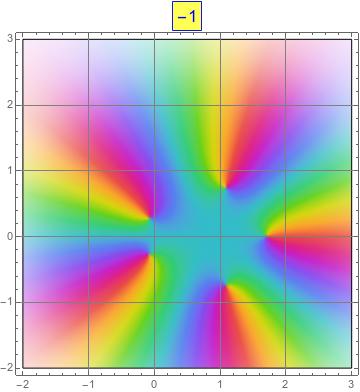}}}	
	\centerline{
	\fbox{\includegraphics[width=2.5cm]{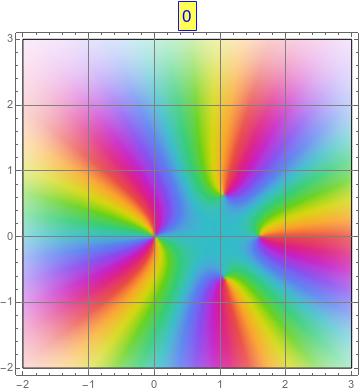}}
	\fbox{\includegraphics[width=2.5cm]{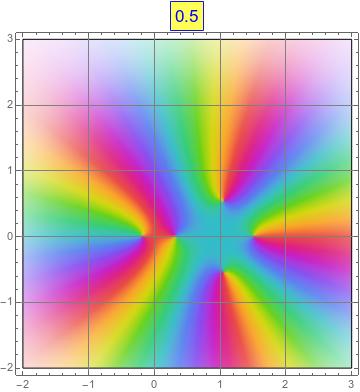}}
	\fbox{\includegraphics[width=2.5cm]{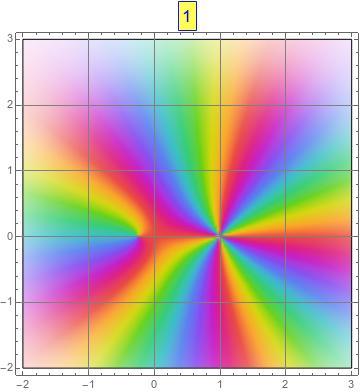}}}
	\centerline{
	\fbox{\includegraphics[width=2.5cm]{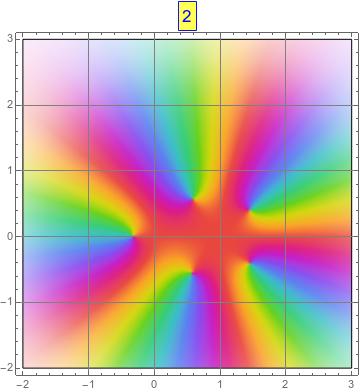}}
	\fbox{\includegraphics[width=2.5cm]{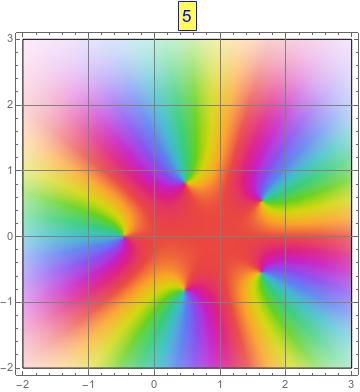}}
	\fbox{\includegraphics[width=2.5cm]{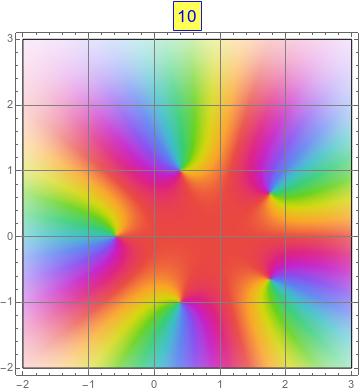}}}
	\centerline{
	\fbox{\includegraphics[width=2.5cm]{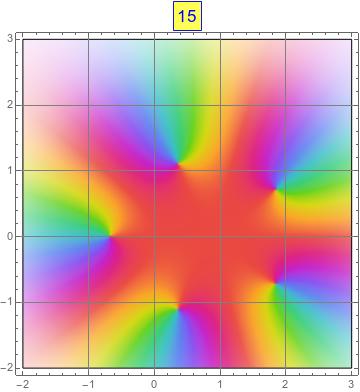}}}
	\caption{{\it 'Dance of the roots', for $a=1$ and $c$ varies from $-15..15$ (from top left to bottom right). The dots represents the zeros and the colors {\bf arg}(z), varying as it goes from $-\pi$ to $\pi$. The first and last one are rotated configurations about $180^o$. }} 
	\label{quintic}
\end{figure}
{\bf 5.} {\it The zero's of the quintic can be found by finding the 'torsion' point of an elliptic curve. The most interesting correspondence occurs when the discriminant of the quintic is zero, i. e. singular curves. In Fig. (\ref{elliptic}) we plotted this case. Specially the double periodic elliptic curves are interesting, because the icosahedron has  two rotations, i.e. over $2\pi/5$ and $\pi$. In Fig. (\ref{quintic}) we also plotted the location of the zeros for $a=1$ and $c$ running from $-15$ to $15$. The parameter $c$ is obviously related to the topology of the icosahedron. }\\

{\bf 6} {\it There is a relation with the Riemann zeta function $\zeta(z)$ and the 'dance' of the zeros of the quintic. The non-trivial zeros of the Riemann zeta function have a geometrical interpretation \cite{aneva2008}. The eigenfunctions of the Laplace-Beltrami operator are automorphic with respect to a discrete group of fractional M\"obius transformations. The Riemann hypothesis encompasses that a meromorphic function $\zeta$  shows fluctuations of spacing between the zeros and can be seen as a 'spectrum' with energy levels $E_n$ with $\zeta(1/2+iE_n)=0$ of a quantum Hermitian operator. One can  construct a quantization condition, generating Riemann zeros with the antipodal geometric application, using conformal symmetry of the hyperbolic (or elliptic in our case) dynamical system and the finite dihedral group $D_4$. 
The boundary conditions for the automorphic functions is compatible with the topology of the projective plane $\mathbb{R}P^2\sim \mathbb{Z}_2$.
Due to the twist, the boundary conditions can serve as quantization conditions, generating a discrete spectrum of the Hamiltonian 
\begin{equation}
H=\frac{1}{2}(qp+pq)=-i\hslash\Bigl(q\partial_q+\frac{1}{2}\Bigr)
\end{equation}
The same route can be applied for our $S^3$ using the Klein construction in 5D.
However, the difference is that in the original treatment,  the antipodal map $(q,p)\rightarrow (-q,-p)$ can be performed by two homotopy classes, depending on whether the final point is reached via a product of even or odd number of reflections on $S^2$. The determinant is $\pm 1$, i.e. reversing or preserving the orientation.
In our model the group of order two of the covering space is replaced by the $\mathbb{Z}_2$ symmetry of the bulk space.}

\section{The new topology and $S^3$ }\label{3}
\subsection{Summary of earlier treatments}\label{3-1}
One can treat the black hole paradoxes, which one encounters in the quantum area, by a revision of the topology. We chose for the antipodal boundary condition, investigated by several researchers\cite{sanchez1987,gibbons1986,gibbons1993,betzios2021,
betzios2021b,thooft2016,thooft2021}.
This approach differs significantly from other approaches to attack the paradoxes, i.e. the information-, the firewall and the complementarity paradox and last but not least, the unitarity of the S-matrix.
When fields are quantized on a manifold, one needs a one-to-one map between the entire asymptotic domain of the maximally extended Penrose diagram.
\begin{figure}[h]
	\centerline{
	\fbox{\includegraphics[width=5.cm]{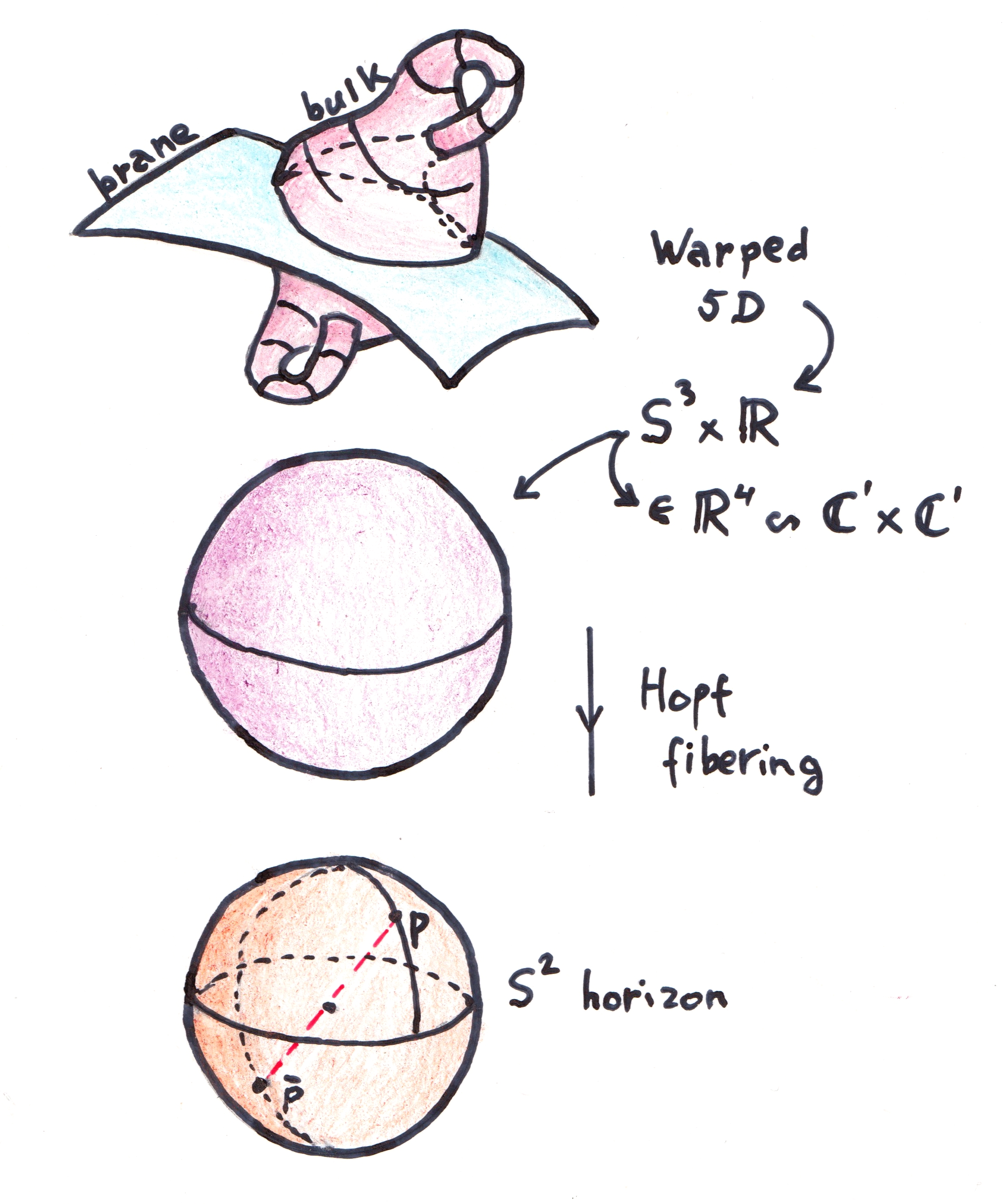}}}	
	\caption{{\it Fibering the $S^3$.}} 
	\label{torusklein}
\end{figure}
In order to maintain CPT invariance by utilizing the spherical harmonics when calculating the commutation relations, one encounters some obstructions \cite{betzios2021b,thooft2021}.  We will see that in our model they disappear.

We consider a Hilbert space $\ket{{\cal Z}}$ on  $\mathbb{C}^1\times \mathbb{C}^1$ as a real vector space $\mathbb{R}^4$ with components  ${\cal Z}=[{\cal V},{\cal W}]$ and ${\cal V}=x+iy, {\cal W}=z+iy_5$. The normalization condition is $|{\cal V}|^2+|{\cal W}|^2=x^2+y^2+z^2+y_5^2=r^2+R^2$, i.e. the $S^3\subset \mathbb{R}^4$.
We write the 3-sphere $S^3=\{x,y,z,y_5\in \mathbb{R}^4|x^2+y^2+z^2+y_5^2=r^2+R^2\}=\{({\cal V},{\cal W})\in \mathbb{C}^2|{\cal V}\bar {\cal V}+{\cal W}\bar {\cal W}=r^2+R^2\}$.
See figure (\ref{torusklein}) and the diagram of Eq. (\ref{3-1}).
We have used the embedding of the Klein bottle in the warped 5D spacetime in our model. To get back to our black hole in the effective 4D spacetime, we can use the Hopf fibration of $S^3\sim \mathbb{C}^1\otimes\mathbb{C}^1$ and apply the antipodal map \cite{thooft2016} to get the $S^2$ of our black hole horizon. 
The Hawking particle created near the horizon at $P(U,V,z,\varphi,y_5)$, travels  on the Klein surface one or two times and leaves the horizon at the antipode $\tilde P(-U,-V,-z,\varphi +\pi,-y_5)$ after a finite proper time. 

\begin{eqnarray}
\begin{tikzcd}
& ({\cal V}, {\cal W})\:\:\:\:\: \in \hspace{-1.5cm}  \arrow[d, "H"]
& \mathbb{C}^2/\{0\} \arrow[d, "\mathbb{C}P^1"]
& S^3 \arrow[l]\arrow[d, "p" ]\\
& \:\:\:\:\:u\equiv\frac{{\cal V}}{{\cal W}} \:\:\:\:\:\: \:\: \in\hspace{-1cm}
&\:\:\:\: \mathbb{C}\cup\{\infty\}\:\:\:\:  =\hspace{-1cm}
& S^2
\end{tikzcd}\label{3-1}
\end{eqnarray}
There are two basic compact self-dual manifolds,  the round metric on $S^4$  and the Fubini-Study metric on $\mathbb{C}P^2$. In our situation we are mainly interested in the quantization on the latter. It is possible to construct a complex two-dimensional Hilbert space ${\cal H}$. The density matrix is $\rho=\ket{{\cal Z}}\bra{{\cal Z}}$ and we can  construct a quantization of $\mathbb{C}P^2$ (\cite{slagter2023,ur1990}.
On the non-orientable Klein surface, meromorphic functions remain constant and our quintic is meromorphic. It is also homeomorphic to the connected sum of two projected planes.
It is also possible to construct geodesics and a minimum surface needed to follow the particles. You can use the torus, on which the orbits are easy to find. Note that the orbit is still in 4D space. However, the geodesics on the Klein surface will never intersect each other because of the non-orientability.
\subsection{Treatment of the Hawking particles}\label{3-2}
We have taken a slightly different approach to the antipodal map, in order to overcome some obstructions, as already mentioned before. To revise the information, firewall and complementarity paradoxes, 't Hooft used the antipodal map on $S^2$, where the non-orientability is provided by the M\"obius strip. His method would restore quantum purity and CPT invariance.  The particles remain entangled. The incoming particles return at the antipode and the particles remain entangled. There is no inside of the black hole, i.e. region II is not 'really there' for the observer inside by the cut-and-paste procedure. 
\begin{figure}[h]
	\centerline{ 
	\fbox{\includegraphics[width=6.cm]{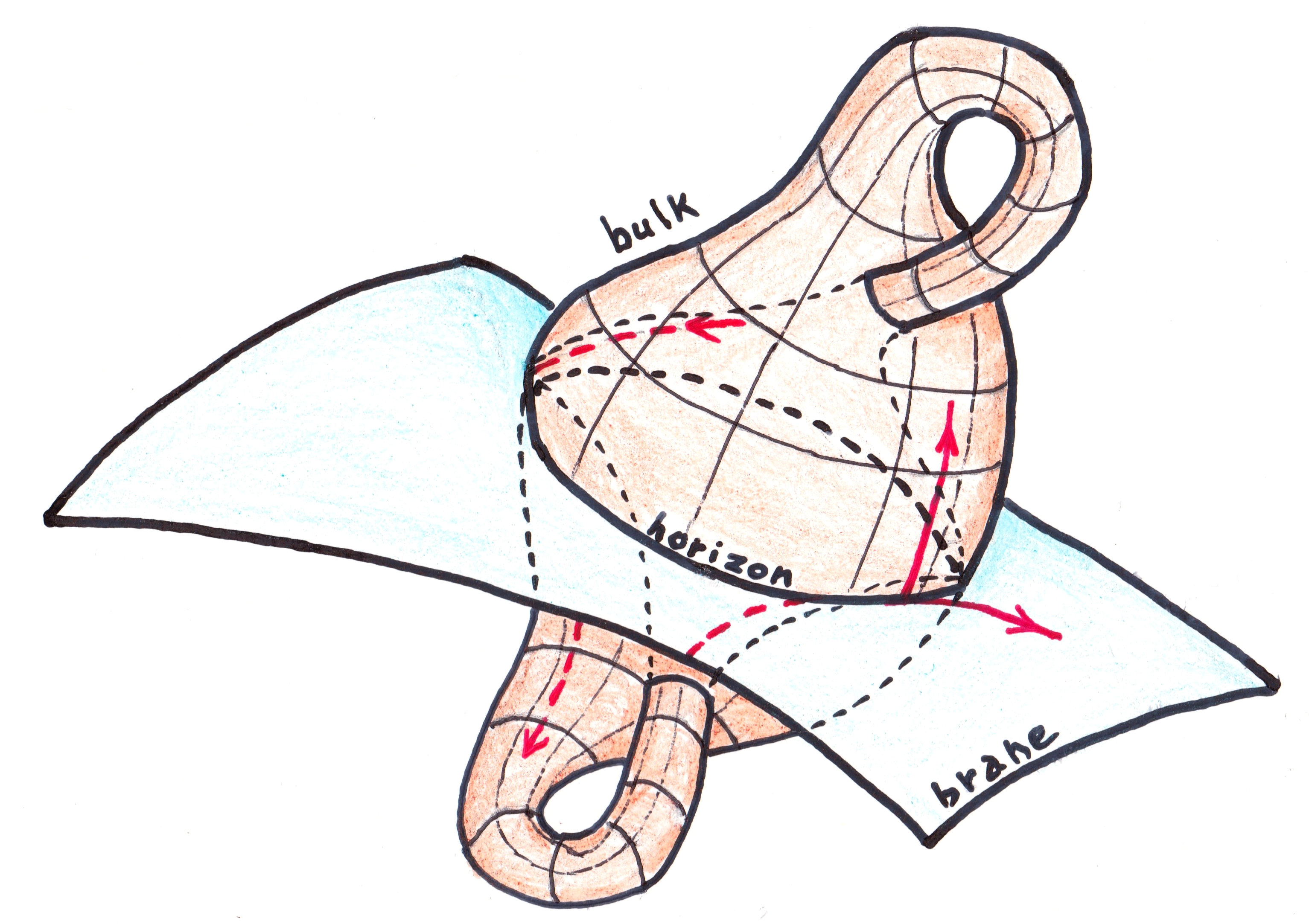}}}	
	\caption{{\it The $\mathbb{Z}_2$ symmetry for the bulk represented as Klein surfaces, with the Hawking particles.}} 
	\label{HawRad}
\end{figure}
In our model we use the Klein surface $\mathbb{K}$.
Our conjecture is that our solution can be seen as a dynamical solution embedded in our $(4+1)$ dimensional spacetime. More precisely, if ${\cal M}$ is the sub-manifold with metric $g$, i.e. the metric on our effective 4D spacetime and ${\cal N}$ the 5D manifold, then ${\cal M}$ is diffeomorphic to the hyperbolic 5D spacetime: $\chi :\mathbb{R}^5\rightarrow {\cal N}$.
The Riemannian manifold $({\cal M},g)$ is conformally flat.
Now physicists are interested in the topology of moduli spaces of self-dual connections on vector bundles over Riemannian manifolds. One reason was that on these spaces the instanton approximation to the Green functions of Euclidean quantum gravity can be expressed in terms of integrals over moduli spaces. One then needs the metric and volume form of the moduli spaces.
We suspect that we can consider $\mathbb{K}$ as a 4-sphere in our hyperbolic pseudo-Riemannian spacetime.
A note must be made about the $\mathbb{Z}_2$ symmetry in the original description of the antipodal mapping \cite{thooft2021}. 
At the boundary of regions I and II in the Penrose diagram, the antipodes were glued together on a 3-sphere, and the transverse $(\theta,\varphi)$ part is a projected 2-sphere. 
In our model it is replaced by the projected 3-sphere $(z,\varphi ,y_5)$ using the $\mathbb{Z}_2$ symmetry of the bulk space $(U,V,z,\varphi,y_5)$. We 'blow up' the 4-manifold to 5D, to handle the singularities in the curvature and to apply the antipodal map.

One can mathematically formulate the topology of a 4-manifold using self-dual connections over de Riemannian $S^4$ \cite{freed1984}. It depends only on the conformal class of the Riemannian metric. This self-dual connection can be interpreted by the conformal map: $\mathbb{R}^4\rightarrow S^4/\{0\}$ as a self-dual connection or instanton. We interpret the Riemannian 5D warped spacetime, an open 5-ball, as an instanton on $\mathbb{R}^4$. 
In the pseudo-Riemannian spacetime, the boundary is the non-orientable Klein surface, which we used for the antipodicity to preserve the pure states of the Hawking particles. It is known that a static solution in $m$ space dimensions is completely equivalent to an instanton in $m$ space-time dimensions.

Furthermore, the evolution of the wave function of the incoming particle must be unitary, i.e. it must satisfy the Schr\"odinger equation $\ket{\chi(t_1)}=U(t_1,t_2)\ket{\chi(t_2)}$ and must be bijective. It is thought that information about the quantum state is preserved during black hole evaporation. The loss of information is incompatible with unitarity.
Thus, 't Hooft's method resolves the controversy between the pure quantum state of the incoming particles and the mixed state property of the Hawking radiation.
The information is retrieved by the outgoing particles. They carry the information of the incoming particles.
In quantum mechanical terms, the incoming position of the wave packet is imprinted on the outgoing momentum. The positions are then highly de-localised. This is how long-wavelength Hawking particles are produced from short-wavelength waves.
In our model, the Hawking particles don't jump infinitely fast to the antipode, but stay on the Klein surface for a while. Moreover, the complementarity in the different notion of the dilaton $\omega$ used by the inside and outside observer is formulated by the invariance of Eq.(\ref{1-1}).

Then we have the problem of handing over the hard gravitons. 't Hooft uses the Shapiro delay method to describe the gravitational interaction near the horizon. Hard particles become soft by taking into account the equations in the transverse spherical harmonics ${\bf Y}_{l,m}(\theta,\varphi)$. The firewall decays due to the identification $(\theta,\varphi)\rightarrow (\pi-\theta,\varphi+\pi)$. The dynamics is factorized in $(l,m)$. 

In our case, the dilaton is written as
\begin{equation}
\omega_{n,m}=\omega(t,r) e^{in\varphi}{\bf Y}_m(\varphi)\label{5-1}
\end{equation}
where  ${\bf Y}_m$  the cylindrical harmonics. It is the general form of the expression ${\bf J}(t,r)e^{in\varphi}{\bf Y}_m(\varphi)$, with ${\bf J}$ the elliptic Jacobi function.
Further, the shape of the probability function makes the gravitons hard in the bulk \cite{ran1999a} and soft on the brane. The presence of the bulk causes the KK contribution of the Weyl tensor on the brane, leading to our new solution.
The spherical harmonics are replaced by the cylindrical ${\bf Y}_m(\varphi)$. The second quantum number is provided by the winding number, because the dilaton behaves as an ordinary quantum field with axial symmetry, just like a scalar field. Furthermore, we obtained PDE's in $(r,t)$ with exact solutions.
There will be a time delay between the incoming and outgoing waves or particles depending on $y_0$, which can be expressed in the parameters of our model. Thus, experimental data could provide an estimate of the bulk dimension.
't Hooft's method requires high wave numbers when approaching the Planck region, and the cut-off is then determined by the number of microstates. In our model we don't need a cut-off. We have the elapsed time on the Klein surface in the bulk, which delivers the degree of 'hardness'.
In our model the wavelength increases with time. This would imply that the particles gain momentum and the quanta become blue-shifted. However, our process takes place in the bulk, so it is not visible from the brane.
One can also apply Schr\"odinger's original treatment of the Klein-Gordon equation on a hypersurface \cite{schrod1957}. He finds conditions for the values of the quantum numbers $(l,m)$\footnote{Schr\"odinger applies the case of an expanding universe.} when the lines of constant phase on the hypersurface will close. In our case we can have the situation where the phase lines spiral around the Klein surface many times without the possibility of intersecting.
It should be noted that the antipodal map is obligatory in order to make the mapping of the coordinates in regions I and II one-to-one, and yet they are distinct because they are spatially separated. There is no summation over 'invisible' quantum states to form the density matrix. We still have pure states.
If you sum over the other states, which are far away in the antipodal region and unobservable to the local observer, you would conclude that these particles have the Hawking temperature. But the state is not thermal at all.
The Hawking particles on one hemisphere are entangled with the antipode.
To the outside observer, the Hartle-Hawking wave function is no longer a thermally mixed state, but a single pure state.  
In general, the antipodes are completely entangled. From the outside, the black hole is not in a stationary Hartle-Hawking quantum state. In other words, there is no complete heat bath. 
Finally, we can study the surface gravity of our model. In the $\varphi^*$ coordinate it becomes
\begin{equation}
\kappa=\frac{2}{5}\Bigl(\frac{10C_2(t-b)^3+6r^2+10a^2-15ar-\frac{5C+a^5}{r^3}}{C_2(t-b)^4+C_3}\Bigr)
\end{equation}
Our solution lookes like  an 'extremal' black hole, i.e. the surface gravity is zero, for 
\begin{equation}
t=b+\frac{1}{a}\sqrt[3]{\frac{C}{2C_2}}\label{5-2}
\end{equation} 
In general, the black hole never becomes extremal in $(t,r,\varphi)$ coordinates.

\subsection{The new antipodal map and CPT invariance}\label{3-3}
We explained in the foregoing sections that in our 5D model the topological changes are different with respect to the standard antipodal application.
We already remarked that there is no causality problem: the particles stay  on the Klein surface for a while. Moreover, we have no cusp singularity, because of the mirror symmetry in the bulk dimension. No closed timelike curves will appear.
Let us summarize again the new topology of the 5D warped spacetime using the Klein surface. 
In the case of the Schwarzschild spacetime in Kruskal-Szekeres coordinates, one could write $(u^+,u^-,\theta,\varphi)\rightarrow (-u^+,-u^-,\pi-\theta,\varphi+\pi)$, which resulted in an obstruction \cite{thooft2021}. The coordinates $(u^+,u^-)$, where $u^\pm\sim  (x,y)$, are given by
$xy=(r/2GM-1)e^{r/2GM}, y/x=e^{t/2GM}$. The time parameter appears in the exponent.
Our antipodal map  was in polar coordinates $(\tau,\rho,\varphi,z,y_5)\rightarrow (-\tau,-\rho,\varphi+n\pi,-z,-y_5$, where we  can use, for example, the twofold change $n=2$ on the Klein surfaces. One has the possibility that the particle reappears after one loop on a Klein surface, or  two loops, after a successive loop on the $\mathbb{Z}_2$ copy.
The sign change in $z$ causes no problem, it is the axis of rotation of the black hole. 
We also mentioned that in our case, we could use the cylindrical harmonics $\omega,\Phi\sim {\bf J}(t,\rho){\bf Y}_m(\varphi) e^{in\varphi}$. 
The same holds for the expression of $u^\pm$ and $p^\pm$.
In the spherical symmetric Schwarzschild case,  $u^\pm$ changes sign from region I to region II and  we have $u^\pm_{l,m}\rightarrow (-1)^l u^\pm_{l,m}$. This is  not the case for our model. We have the freedom $(-1)^{m+n}X(t,\rho)e^{in\varphi}{\bf Y}_m$.
Moreover, the commutation rules for $u^\pm$ will not cause the obstruction.

We know that a quantum Hamiltonian captures the near horizon dynamics of S-matrix for a black hole \cite{betzios2021}. One should also have CPT symmetry, at least as a local symmetry, obtained by symmetry breaking from a global discrete spacetime symmetry. The latter requires the specification of a choice of time, i.e. a choice of gauge, because there are no global symmetries in quantim gravity. Our new antipodal  boundary condition, relates points across the two sides of the Penrose diagram. We also found  that $UV=(r-r_H)^{\kappa r_H/(r_H+a)^3}$ is invariant under $U\rightarrow -U, V\rightarrow -V$. The $(U,V)$ basis can  be used to describe  the outgoing and in ingoing modes of the wave functions. The $(U,V)$ are conjugated variables.
One can perform the calculations in the  parameter $M_{Pl}/M_{BH}$ on the state formed between the times later then the time of formation  and earlier than the time of evaporation \cite{betzios2021,betzios2021b}. But what about the situation just at the formation time? Our exact model can handle this situation.

The initial obstruction occurs due to impossibility to formulate the commutation relations between $U$ and $V$, where the time dependency is a priori defined by $e^{t/2GM}$ \cite{thooft2021,betzios2021}. The contradiction occurs upon the rotation invariance to the antipodal points, when applying the expansion in spherical harmonics $Y_{lm}(\Omega)$. The even $l$ modes are then forbidden. This resulted in a single spacetime exterior. 
The obstruction  could be avoided by the right interpretation of the central causal diamond  $UV\sim \hslash$ for the wave functions. Points in the $(U,V)$ plane close to the origin, are not sharply defined due to Heisenberg's uncertainty relation. So one can define a set of discrete conformal inversions by the M\"obius transformations.
It turns out \cite{betzios2021b,aneva2008} that the transformations generate a finite group, i.e. the dihedral group $D_4$ with eight elements. One uses that $\Pi: \mathbb{R}P^2\cong \mathbb{Z}_2$. The result is that CPT is maintained on the complete wave functions, by using two $S^2$ with complementary antipodal identifications. 
This is just what we found in our 5D model, where the dihedral group $D_4$ was replaced by the $D_5$ and the double complementary Klein surfaces in one dimension higher.

One can apply a complex two level Hilbert space \cite{ur1990}. They form the state vectors.  Pure states live on the Bloch sphere $S^3$, where antipodal points form a qubit state. Mixed states are located in the interior. Running point on $S^3\in \mathbb{R}^4$ are related to its stereographic images on the equatorial plane $\mathbb{R}P^3$. The map is non-orientable and conformal, as required.
In order to apply this model on the warped 5D spacetime, we need the  Klein surface \cite{slagter2024,slagter2025}.
A striking feature of the 3-sphere is that it also contains Killing vector
fields, in contrast to the 2-sphere. Every vector field on $S^2$ must have a fixed point somewhere, however, not  for the 3-sphere.
It is noteworthy that all streamlines of the Killing vector field $\xi$ are also geodesics. 
One can make the connection with a 4-dimensional dodecahedron and icosahedron (already  noted by Poincaré). This 'Poincar\'e ball' can be seen as a dodecahedron space where we use $[e^{2\pi i /10}]^5=-1$. The group $D_5$ describes the symmetries. The space is a manifold, because it is locally homeomorphic to Euclidean space, even in the vicinity of the vertices and points on the edges. There is another advantage. We can apply easily the spinor mathematics for fermions. They are the carriers of the representation of the $SL(2,\mathbb{C})$. It was Dirac who presented a visualization by his 'dish' model (or 'scissors' model). In mathematical language, the $SO(3)$ of rotations in a 3-space is double covered by the group of 3-spheres of quaternions in a 4-space. The configuration space is denoted as $\mathbb{R}^3\times SO(3)$. What Dirac tells us is that a loop in $\mathbb{R}^3\times SO(3)$ that corresponds to a rotation of $2\pi$ is not homotopically trivial, but that the loop that corresponds to a rotation by $4\pi$ is homotopic to the trivial loop.
Every rotation of $\mathbb{R}^3$ can be uniquely specified
by a rotation axis, an angle, and a direction of rotation around the axis. All this information can be encoded in a single object, namely a
vector $\vec{n}$ in $\mathbb{R}^3$ of at most magnitude $\pi$. The rotation axis is then the line along $\vec{n}$, the rotation angle is $|\vec{n}|$ and the direction is determined by the right-hand rule. Note that a rotation along  $\vec{n}$ by an angle $\theta$ with $\pi \leq \theta \leq 2\pi$ is equal to a
rotation along $-\vec{n}$ over $2\pi-\theta$, so the restriction on $|\vec{n}|$ is necessary to ensure that the correspondence between rotations and vectors is 1-to-1. The set of vectors $|\vec{n}|$ in $\mathbb{R}^3$ with $|\vec{n}| \leq \pi$ is simply the closed sphere with radius $\pi$ around the origin. However, a rotation about $|\vec{n}|$ by $\pi$ is the same as a rotation about $-|\vec{n}|$ by $\pi$. So antipodal points on the boundary of this sphere represent the same
rotation and must therefore be identified, in order that this correspondence
with rotations is bijective.
Performing this identification yields the real projective 3-space (topologically, the radius of the sphere is of course irrelevant). 
We can write out the one-to-one correspondence we just described geometrically to show that it is in fact continuous as a mapping from $\mathbb{R}P^3$ to $S O(3)$ . Further, $\mathbb{R}P^3\sim S^3$, so $\Pi : \mathbb{R}P^3\cong \mathbb{Z}_2$.
Because $\mathbb{R}P^3$ is compact and a continuous image of $S^3$ , we find that $SO(3)$ is homeomorphic to $\mathbb{R}P^3$ . In particular, $\Pi:  SO(3)\cong \Pi: \mathbb{R}P^3\cong \mathbb{Z}_2$. So $\Pi : \mathbb{R}^3\times SO(3)\cong\Pi :(\mathbb{R}^3\times \Pi :SO(3))\cong \{0\}\times \mathbb{Z}_2\cong \mathbb{Z}_2$.
In order to get back to the $S^2$, we applied the Hopf fibering (Fig. \ref{torusklein}. The real or complex projective plane $\mathbb{R}P^n$ and $\mathbb{C}P^n$ in $n$ dimensions can be created by a set of 'lines' through the origin in $\mathbb{R}^{n+1}$ and $\mathbb{C}^{n+1}$ respectively. Remember that $S^3$ can be seen as unit sphere in $\mathbb{C}^1\times \mathbb{C}^1$. The Hopf map $(u,v)\mapsto  u/v$ is then a map $S^3\sim S^2$, with generators of the dihedral group.
Finally, we note once again that our time symmetry  follows directly from the PDE's (both on the pseudo-Riemannian and Riemannian spacetimes) and is not generated by a separated Hamiltonian. 
So we can avoid the assumption that region II in the extended Penrose diagram is the interior which contains nothing but quantum clones of the real physical variables. Moreover, one maintains unitarity directly for the in and out particles.
\section{The instanton}\label{sec4}
The notion of an instanton, or pseudoparticle, was introduced by 't Hooft and represents a localized finite action topological invariant in field theory. It was first developed in Yang-Mills theory, in particular in $SU(n)$ gauge theories.
The classical field equations with self-dual field strength then gave rise to a new non-perturbative treatment. The Minkowski spacetime was replaced by a Wick rotation $t\rightarrow i\tau$, where $\tau$ is a special parameterization of the favoured tunneling path through the configuration space. Feynman's path-integral approach to quantization is then unambiguous. Solutions are concentrated around a point in four-dimensional Euclidean space, and at one instant there is a 'bump' in the field strength, which quickly dies away. 
It turns out that self-duality in the Yang-Mills model, leads to first-order differential equations. They played a fundamental role in non-abelian gauge theories. See, for example, the textbook of Felsager \cite{felsager1998}.

There are striking similarities between these instantons and their gravitational counterparts. They are obtained in general relativity by a Wick rotation on Riemannian spacetime. The field equations then lead to first order differential equations \cite{gibb1993}.

It is now known that there are Euclidean self-dual instanton solutions in gravity theory. The self-duality is then expressed in terms of the Riemann curvature tensor. They are thought to play a role in understanding quantum gravitational effects near the horizon of BH's, such as the evaporation process and the formation of Planck-sized BH's\footnote{Recently, primordial BHs have been observed, which could not be formed by contracting matter.}.
Well-known examples are the Eguchi-Hanson (EH) solution \cite{eguchi1978} and the Fubini-Study solution, which can be described on a $(\mathbb{C}^1\times\mathbb{C}^1)$ K\"ahler manifold \cite{joyce1995}.

\begin{figure}[h]
	\centerline{
	\fbox{\includegraphics[width=5.cm]{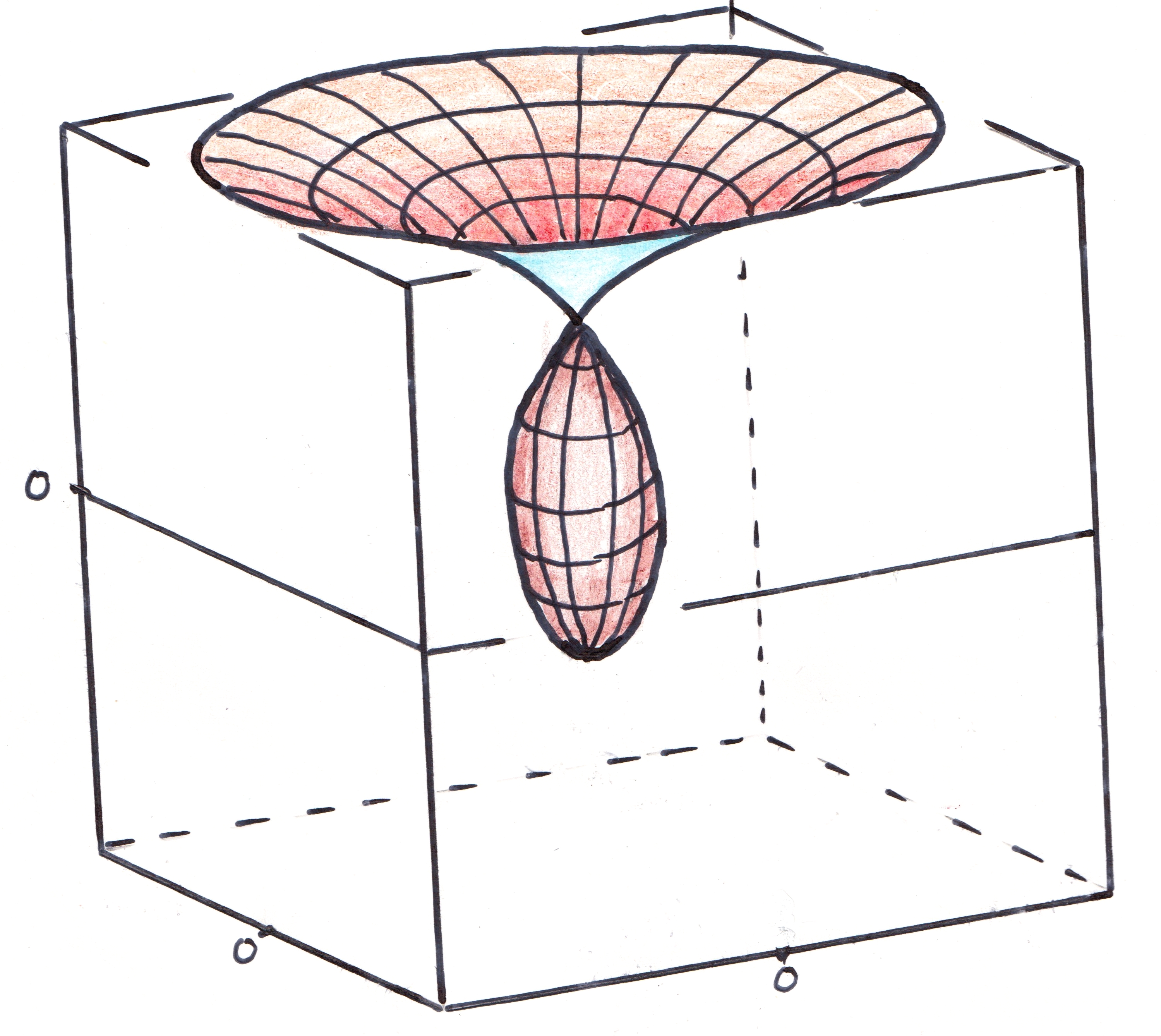}}}	
	\caption{{\it Shape of the inside of the black hole, plotted as surface of revolution. $-\frac{a}{4}<r<a$, with $a$ the horizon. There is no $r=0$ singularity.  }} 
	\label{instanton}
\end{figure}

Let us return to our solution. If we switch to Riemannian space with $t\rightarrow i\tau$, we get the same solution as in Eq.(\ref{2-7}) and (\ref{2-8}).
The line element can be written in the stationary case and for $C_1=0, k=3$ as
\begin{eqnarray}
ds^2=\omega^{4/3}\bar\omega^2\Bigl[\frac{(a+4r)(r-a)^4}{5r^2}d\tau^2+dr^{*2}\cr+dz^2+r^2d\varphi^{*2}+dy_5^2\Bigr]\label{4-1}
\end{eqnarray}
with
\begin{eqnarray}
r^*=\frac{1}{5\sqrt{a}}\Bigl[ 6 arctanh\Bigl(\sqrt{\frac{a+4r}{5a}}\Bigr)
+\frac{\sqrt{5a(a+4r)}}{r-a}\Bigr]\label{4-2}
\end{eqnarray}
and $-\frac{a}{4}<r<a$. It  describes the interior of the black hole. In Fig.\ref{instanton} we have plotted $r^*$, valid only inside the horizon $r_h=1$, for $a=1$.
We will also change the azimuthal variable by $d\varphi^*=d\varphi +N^\varphi dt$.
The region $r<\frac{a}{4}$ is inaccessible. It would be challenging to consider large values of $k$. We will then get a horizon and the singularity for $r \rightarrow 0$ will never be reached.
It became quite clear that Euclidean self-dual solutions in Einstein's theory play a role in quantum gravity. These solutions have vanishing classical action and non-trivial topological invariants. Moreover, there is a similarity with the YM Euclidean self-dual solutions \cite{eguchi1978}.
Despite their Euclidean origin, instantons, like solitons, play a fundamental role in quantum theories. Their importance arises, for example, in  tunneling effects using the WKB approximation.
Instanton-like solutions are expected to be all self-dual, localized in Euclidean spacetimes and free of singularities.
They give a dominant contribution to the path integral that is as important as that of the flat metric itself. They also have time-parity inversion.
However, gravitational instantons do not contribute to the zero-frequency modes of a spin-1/2 Dirac particle, while they produce zero-frequency solutions of spin-3/2 particles.
It is conjectured (\cite{gibb1993}, \cite{atiyah1978}) that there is a deep connection between the gravitational instanton and the alternating discrete symmetry group ${\cal A}_5$ which serves as a discrete subgroup of $SO(4)$ of its boundary.
{\it A gravitational instanton is a four-dimensional regular spacetime $M$, self-dual Ricci curvature and asymptotically local Euclidean, whose boundary  has a ${\cal A}_k,D_k, T, 0, I$ symmetry}\footnote{These are the dihedral, M\"obius, tetrahedral, octahedral and icosahedral groups respectively.}. In fact, every finite subgroup of $S^3\sim SU(2)$, is either cyclic or conjugate to  these binary subgroups.
For our 5D warped spacetime, our boundary is the Klein surface, embedded in $\mathbb{R}^4\sim \mathbb{C}^1\times \mathbb{C}^1$ and the symmetry the double cover group  $I$ of the  icodahedral.
An interesting study has also been done by Gibbons \cite{gibb1993}. He studied, on $\mathbb{C}P^2$,  a gravitational instanton surrounded by an event horizon and compared the solution with the $SU(2)$ Yang-Mills instanton. 
This study is based on the Riemannian solution of the Einstein equations with a cosmological constant on $\mathbb{C}P^2$ by applying the Hopf-fibering. The Weyl tensor is then anti-self dual and they conclude that there is no Lorentzian counterpart connection, i. e. it cannot be related to a real 4-dimensional manifold. Our model can. 

Finally, we comment on the notion of a quantum black hole, compared with a hydrogen atom.  It is a bold statement \cite{bekenstein1974}.  In atomic physics, emission spectra show a hierarchy of level splitting caused by the breaking of different symmetries. For the hydrogen the $O(4)$ symmetry of the Coulomb problem is broken, resulting in the fine structure splitting of the Rydberg-Bohr spectrum, caused by the relativistic spin-orbit interaction  and Thomas precession.
And we have the hyperfine splitting by the Lamb shift and the vacuum polarization effects. The question is if one can compare this viewpoint with the $2^n$-fold degeneracy of the H-atom. Moreover, will there room for considering a kind of symmetry breaking, analogous to the H-atom, which could split the black hole 'lines'? And what about the construction of the operators for the black hole, $(\hat M, \hat A, \hat Q, \hat G. \hat {\bf J}, \hat J_z)$, i.e the mass, area, charge, magnetic charge  angular momentum respectively.

But eigenstates of the black hole is of a different quantum mechanical order.
However, we found in our model a remarkable discretization of the black hole's 'eigenstates' of the distribution of the zero's, stereographically projected on $S^3$. This subject is currently under study by the author.
\section{Conclusions}\label{summ}
We discussed the exact solution of a conformally invariant Randall-Sundrum 5D warped Kerr-like spacetime. 
We applied the topological antipodal  boundary condition for the 3-sphere.
The vacuum solution found on the conformal invariant 5D warped spacetime, where the metric is determined by a quintic polynomial, can be described by a Laplace-type equation on a Klein surface. The zeros of this quintic polynomial depend on the  constraints on the parameters of the model under consideration.
In a special case, the zeros can be expressed  analytically  as $\sim (r-r_H)^{k+1}[r(k+1)+r_H]+C(k+2)$ where $k$ is an integer and $C$ a constant. They are related to the eigenvalues of the Laplacian on the Klein surface.

The antipodal mapping  of regions I and II in the Penrose diagram is accompanied by the continuous transition in the extra bulk dimension by means of the non-orientable Klein surface embedded in $\mathbb{R}^4\sim \mathbb{C}^1\times \mathbb{C}^1$ by means of a Hopf fibration. Measuring the elapsed time that the Hawking particles spend on the Klein surface could give an estimate of the dimension of the bulk.

The vacuum solution describes an instanton in Riemannian space. 
The need for a mechanism for the formation of primordial black holes at Planck-scale dimensions is urgently demanded by the discovery of black holes and mature galaxies in the early universe. Instantons could  trigger this mechanism.
In suitable coordinates, the inside of the black hole shows no central singularity, by applying Cauchy's theorem. 

The information problem could be  solved by the antipodal boundary condition, i.e. by removing the inside of the black hole. The result was that the black hole is not thermal. The reason is that there is no 'hidden sector' anymore to produce the thermodynamically mixed state of the Hawking radiation. All states remain pure for the outside observer.
In our model, the antipodal map on the brane is accompanied by the $\mathbb{Z}_2$ symmetry in the bulk in the Randall-Sundrum model. No 'cut-and-past' is required.  This means that there is no  infinitely fast transport to the other side of the hemisphere. The Hawking particles remain  on the Klein surface for a while, as seen by the outside observer. A local observer will not notice the $r=0$ singularity in a suitably chosen coordinate system for the $S^3$, which is conformal flat. 

There is no need for quantum clones in the extended Penrose diagram, in order to overcome some obstruction in constructing of a quantum version, i.e. quantizing the position and momentum distributions using commutation relations. Unitarity and CPT invariance can be maintained in our axially symmetric Kerr spacetime.
The near horizon region described in   Kruskal–Szekeres coordinates $(U,V)$, with $UV=(r-r_H)^{\alpha r_h/(r_h+a)^3}$. Because $\Delta U \Delta V\geq \hslash \lambda$, there exist minimal size causal diamonds for each harmonic around $U=V=0$.
No forbidden even modes of spherical harmonics applies and the map $(U,V)\rightarrow (-U,-V)$ can be used. No approximation with a Regge-Wheeler potential is needed. 
Moreover, in our model we need cylindrical harmonics without a restriction on the quantum numbers. 
The axial symmetry  becomes manifest by the stereographic projection of the 3-sphere on the hyperplane $\mathbb{R}^3$ given by $y_5=0$.
Our Klein surface  possesses the right antipodal topology, to obtain quantization conditions and to generate a discrete spectrum of the Hamiltonian.
The covering space is the $\mathbb{Z}_2$ symmetry of the bulk space.

A geometric quantization of our Klein surface can be constructed using a two-level Hilbert space on the complex hyper-surface $\mathbb{C}^1\times\mathbb{C}^1$.

A possible relation is stipulated between Beckenstein's quantum black hole and our solution. Using Galois group structure, we observed that the ordered roots of the quintic are related to to points on the icosahedron. Written in the Brioschi form, the 'dance of the roots' is described by a permutation of the roots. The relation with the fractional M\"obius transformations can be  made.
The discrete 'chaotic' spectrum for the black hole microstates could be related to the Riemann zeta functions.


\end{document}